\documentclass[article]{elsarticle}

\usepackage{booktabs}

\usepackage{latexsym,amssymb}
\usepackage{enumerate}
\usepackage{amsmath}
\usepackage{natbib}
\usepackage{rotating}
\usepackage[colorlinks=true,linkcolor=blue,citecolor=blue]{hyperref}
\usepackage{changepage}
\usepackage{scrextend}

\newcommand*{\cf}{\fontfamily{lmtt}\selectfont}

\usepackage[procnames]{listings}
\usepackage{color}
\usepackage{listings}

\definecolor{keywords}{RGB}{255,0,90}
\definecolor{comments}{RGB}{0,0,113}
\definecolor{red}{RGB}{0,0,0}
\definecolor{green}{RGB}{160,150,0}
\definecolor{black}{RGB}{0,0,0}
 
\begin{document}
\begin{frontmatter}

 \title{AstroBlend: An Astrophysical Visualization Package for Blender}

\author{ J.~P.~Naiman}
\address{Harvard-Smithsonian Center for Astrophysics, The Institute for Theory and Computation,
60 Garden Street, Cambridge, MA 02138, USA}

 \begin{abstract} 

 The rapid growth in scale and complexity of both computational and observational astrophysics over the past decade necessitates efficient and intuitive methods for examining and visualizing large datasets.  Here, I present {\it AstroBlend}, an open-source Python library for use within the three dimensional modeling software, {\it Blender}.  While {\it Blender} has been a popular open-source software among animators and visual effects artists, in recent years it has also become a tool for visualizing astrophysical datasets.  {\it AstroBlend} combines the three dimensional capabilities of {\it Blender} with the analysis tools of the widely used astrophysical toolset, {\it yt}, to afford both computational and observational astrophysicists the ability to simultaneously analyze their data and create informative and appealing visualizations.  The introduction of this package includes a description of features, work flow, and various example visualizations.  A website - www.astroblend.com - has been developed which includes tutorials, and a gallery of example images and movies, along with links to downloadable data, three dimensional artistic models, and various other resources.  
\end{abstract}

 \end{frontmatter}

\section{Introduction}

With the growth of multi-dimensional astrophysical datasets from simulations and observations, the need for greater interaction with such complex systems has also increased \citep{barnes2008,turk2011,punzo2015,vogt2015,taylor2015}.  Not only do larger simulations and observation programs necessitate efficient ways to investigate their data  \citep{turk2011,turk2013,punzo2015}, but comparison between the outputs of multiple astrophysical codes and observational surveys requires a cross platform application capable of processing numerous data formats \citep{turk2011,agora2014}. 

The need for multidimensional data analysis is twofold.  Collapsing multidimensional data into two dimensions for the purposes of a plot in a paper generally necessitates a loss of information.  This loss can seek to illuminate the salient physical principles within the data, however this is often not the case \citep{vogt2015}. In addition, the need to efficiently interact with multidimensional data will only grow with new and more complex astrophysical datasets as automatic analysis packages are not always reliable \citep{whiting2012,taylor2013}.

In this new large dataset paradigm, development of both efficient means of visualization along side quantitative analysis is required - both methods of analyzing the data are complementary \citep{goodman2012}.  Recent innovations in the computer graphics and gaming industry can be exploited for this multi-pronged approach to data investigation \citep{taylor2015,kent2015}.  In addition to increasing the rate of information exchange within scientific collaborations, as technological advancements in graphics and gaming progress, innovative methods for discussing scientific concepts with the general public can be developed \citep{vogt2013,steffen2014,vogt2014,brown2014,madura2015}.

In the past several years there has been a growing interest in the use of the three dimensional modeling and special effects software {\it Blender} as a scientific visualization and analysis tool.  {\it Blender} is currently being utilized in fields as diverse as biology in order to study proteins \cite{zini2010} and to visualize planetary surface topology with data from Mars, Earth and the Moon \cite{flor2015}. 
In astronomy, the work of \cite{kent2013,kent2015} serves as an introduction to many of {\it Blender}'s features for the field, and focuses mainly on methods to generate physically accurate artistic models of astrophysical objects.  While \cite{kent2013,kent2015} provides initial methods for viewing observational and simulation data, the {\it FRELLED}\footnote{http://www.rhysy.net/frelled.html} package provides a detailed interface for uploading and interacting with multidimensional observational and simulated datasets \citep{taylor2015}.  {\it FRELLED}  is styled for the extraction of HI sources, which makes use of {\it Blender}'s volume rendering capabilities and includes methods to record source information, decreasing extraction times by more than an order of magnitude.  While initially developed as predominantly a FITS viewer, it provides methods to upload and interact with data from Adaptive Mesh Refinement (AMR) and Smooth Particle Hydrodynamics (SPH) simulations.

In this paper, I introduce {\it AstroBlend}, a Python package which provides complimentary capabilities to these other methods.
Not only does {\it AstroBlend} allow users to compare data from multiple sources (simulations and observations), it is designed to lower the barrier to entry for interaction and visualization of astrophysical datasets in a three dimensional context.  In this first release, there is no support for the volumetric interaction with data as provided with the {\it FRELLED} package, however it is being developed for subsequent releases.
While other visualization tools have been developed for the analysis of large datasets (ParaView \citep{paraview}, VisIt \citep{visit}), they lack easy integration between datasets and physically motivated artistic models.  As a predominately three dimensional modeling software, {\it Blender} provides access to a wealth of artistic representations of scientific concepts.  In addition, by building a {\it yt} \citep{turk2011} backend into {\it Blender}, we allow for both analysis and visualization to be produced in tandem with the support of the full set of tools available in {\it yt}.  The focus of this paper will be on the visualization capabilities of {\it AstroBlend} in {\it Blender}.

The organization of this paper is as follows: after a brief review of the {\it Blender} software and GUI in section \ref{section:blender}, an example workflow including the import of astrophysical data and image rendering is presented in section \ref{section:workflow}.  The different methods for importing datasets are discussed in section \ref{section:inputFormats}.  Options for manipulation of lighting and camera positioning are summarized in sections \ref{section:lighting} and \ref{section:cameraSetup}, respectively.  Several three dimensional annotations are discussed in section \ref{section:annotations}.  Preliminary camera motion options for the creation of movies are shown in section \ref{section:cameraMotions}.  The paper concludes with several example renders and interactive three dimensional models in section \ref{section:examples} and a discussion of future plans in section \ref{section:discussion}.

\section{Blender} \label{section:blender}

{\it Blender} is an open source three dimensional modeling and animation software used by millions\footnote{http://www.blender.org/about/website/statistics/} of visual effects artists, animators and an increasing number of astrophysicists \citep{aragon2010,kent2013,taylor2015,kent2015}.  The software is cross platform - packages for Linux, Windows and Mac computers are provided on the download website\footnote{http://www.blender.org/download/}.  In addition to the rendering of images from three dimensional models, {\it Blender} provides a game engine for interactive graphics along with support for animating two and three dimensional characters.  
Many other tools are present in {\it Blender} including methods for rendering complex surfaces and volumes (with emission, transmission, absorption, reflective and texture components), compositing images, stereoscopic support for three dimensional graphics, and the ability to export models and lighting in a variety of formats for importing into other 3D modeling and gaming software like the Unity\footnote{Unity download and information can be found here: https://unity3d.com/} game engine (Wavefront (OBJ) formatting) or 3D printing software (STereoLithography (STL) formatting).
In addition, {\it Blender} includes some prescriptions for modeling physical phenomena, including fluids using the Lattice Boltzmann Method \citep{chen1998}, smoke as particles based on \cite{monaghan1992}, and rigid body mechanics using Newtonian physics.  Here, we will focus predominately on the rendering of images, specifically from three dimensional astrophysical models and datasets. 

\begin{sidewaysfigure}
\centering
\includegraphics[width=1.0\textwidth]{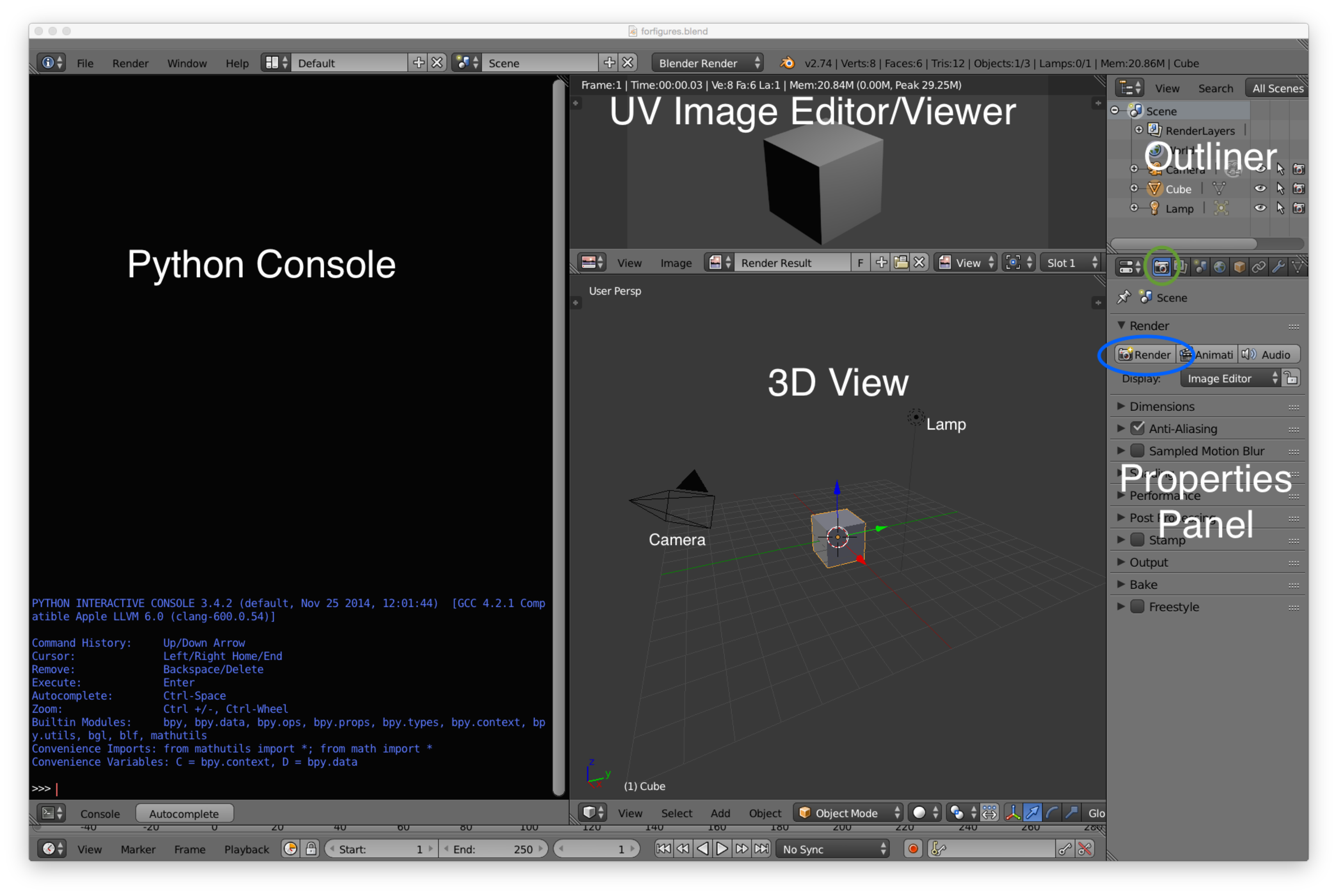}
\caption{A typical window setup for {\it Blender} involves utilizing {\it Blender}'s customizable GUI.  On the left, the Python Console can be utilized to interact with the three dimensional objects and renders 
via the Python scripting language.  The 3D View panel shows the objects in the three dimensional space - in this example, a cube, a camera and a lamp.  These three objects are also 
listed in the Outliner panel which summarizes the objects active in the 3D View panel.  The UV Image Editor (or Viewer) shows the result of a preview render created by pressing the camera button in the Properties Panel (highlighted with a green circle) followed by the Render button (Properties Panel, highlighted with a blue circle).  In addition to rendering buttons the Properties Panel also contains information about objects' materials, textures and other properties.}
\label{fig:usual}
\end{sidewaysfigure}

Rendering an image in {\it Blender} consists of supplying a rendering engine with defined light sources to calculate light transmission, absorption, reflection, and shadows generated by objects placed in {\it Blender}'s three dimensional space.  
An example of an object and a light source placed in the three dimensional space is shown in Figure \ref{fig:usual}.  Here, {\it Blender}'s highly customizable interface has been modified to include a view of the interactive three dimensional space in the 3D View panel, a Python Console, and a UV Image Editor panel.  With this setup, when the {\cf camera} button in the Properties Panel is selected (highlighted with a green circle in Figure \ref{fig:usual}), followed by clicking the {\cf Render} button (highlighted with the blue circle in Figure \ref{fig:usual}), {\it Blender} uses its Internal Renderer to calculate the path of light from the lamp in the 3D View panel (shown as two concentric black circles) as it bounces off the cube and into the camera (shown as the pyramid with a black arrow on top in the 3D View panel).  The resulting image is shown in the UV Image Editor panel of Figure \ref{fig:usual}.  One can interact with the objects in the 3D View panel directly - moving the cube, lamp, or camera with the mouse and re-rendering will generate a different image in the UV Image Editor.
Rendering an image internally without generating an external file can be done through the default Internal Renderer as shown in Figure \ref{fig:usual}, or through the newly developed Cycles Renderer\footnote{Comparison between the Internal and Cycles Renderers can be found here: https://cgcookie.com/image/blender-internal-vs-cycles/}, which creates more photo realistic images than the default internal renderer and allows for both GPU and CPU rendering.   Both the default Internal and Cycles render engines can provide realtime, interactive preview renderings of objects in the 3D View.  In addition to providing render previews either in the UV Image Editor, or in real time interactively, the two internal renders can generate images to be saved as external images in a variety of formats (png, jpg, etc).  While the example shown here uses lamps to light the scene which can often enhance visualizations, more physically realistic lighting from the data itself can be utilized to illuminate the scene as discussed in section \ref{section:workflow}.  

One can also pass data from {\it Blender} to one of the many external renderers such as Mitsuba\footnote{https://www.mitsuba-renderer.org/}, LuxCore\footnote{http://www.luxrender.net/wiki/LuxCore}, and Aqsis\footnote{http://www.aqsis.org/}, among others.  Each of these external renders can produce images of higher quality than {\it Blender}'s default internal renderer, with trade-offs in user interface, speed, and accuracy\footnote{For a comparison of three of the most popular external renderers see https://bartoszstyperek.wordpress.com/2015/02/14/luxcore-vs-mitsuba-vs-cycles-round-3/}.   The images shown in this paper are produced using only {\it Blender}'s Internal Renderer, however Cycles rendering will be supported in a future release of the {\it AstroBlend} library.

{\it AstroBlend} simplifies the importing process of astrophysical datasets and generating of visualization and analysis plots in {\it Blender}. The library makes heavy use of {\it Blender}'s Python API\footnote{https://www.blender.org/api/blender\_python\_api\_2\_76\_2/} to generate scripts which can either be called externally or pasted into the Python Console shown in Figure \ref{fig:usual}. Note that while {\it Blender} is a powerful and widely used tool, it isn't strictly a tool used by astronomers - {\it Blender} is generally used for low resolution models (low polygon-count surfaces) with physical accuracy modeled by ``painting on" textures, while astronomers produce high resolution simulations (resulting in high polygon-count surfaces and particle clouds).  In addition, an animator typically generates one three dimensional model which is then manipulated as the scene is animated, while astronomers instead represent the animation of a scene with stepping through collections of time series simulation files.  This distinction manifests itself in the way we manipulate our data in {\it Blender} - particularly in how objects are loaded and deleted.   These subtleties are outlined below in section \ref{section:workflow}.
Finally, there are many other features in {\it Blender} such as keyframing (an animation technique), and UV unwrapping (a modeling technique) that may be useful in the future, but will not be discussed in the context of this paper (see: \cite{kent2015} for details on these processes).

\section{Example Workflow} \label{section:workflow}

\begin{sidewaysfigure}
\centering
\includegraphics[width=1.0\textwidth]{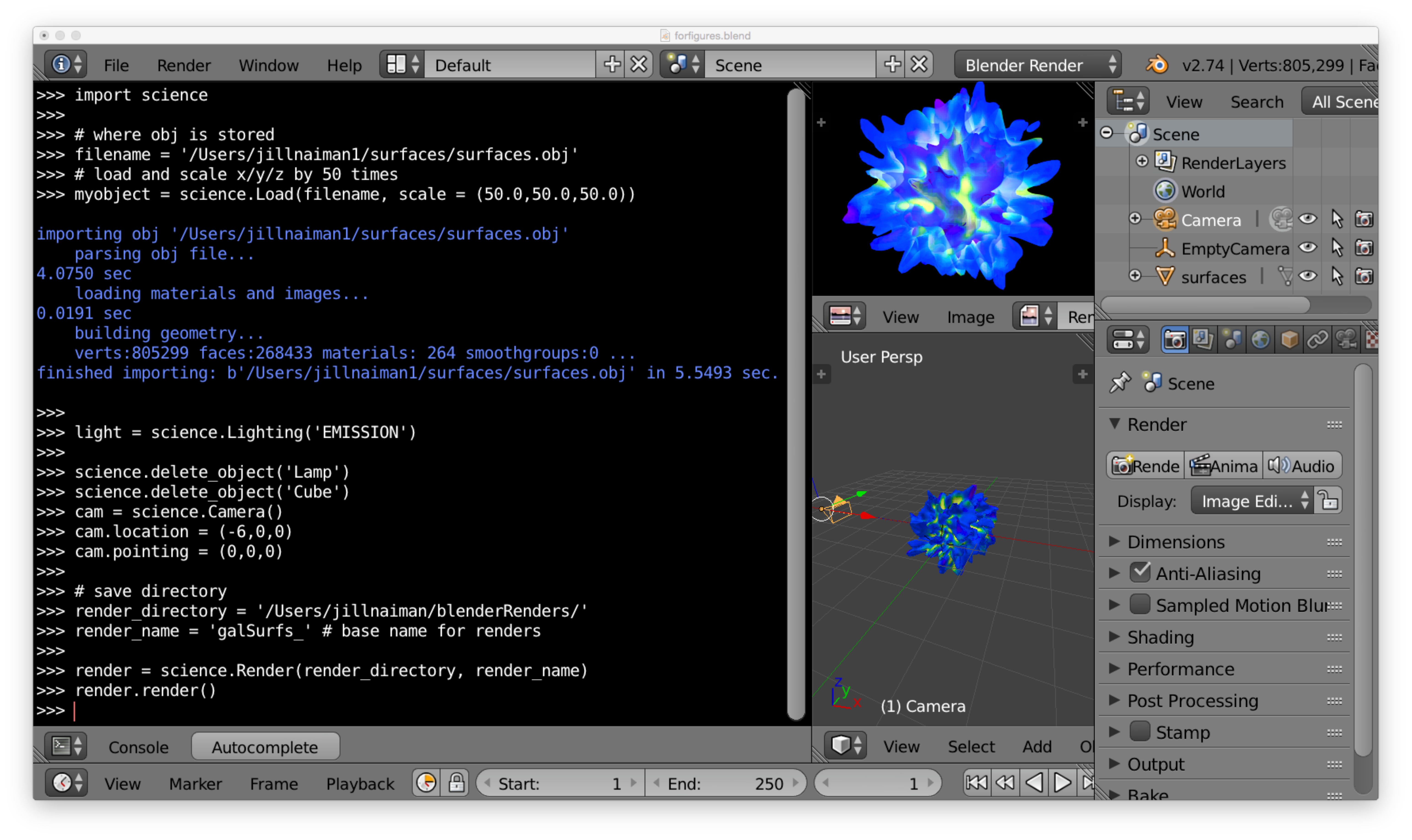} 
\caption{Generating an image from pre-generated set of isodensity surfaces with the isolated galaxy test simulation of \cite{oshea2004}.  The two surfaces are colored by temperature and each face of each surface has an emissivity proportional to its bremsstrahlung emission $\epsilon \propto \rho^2 T^{1/2}$.  In the Python Console (left panel) the code for loading the surfaces (stored in the file surfaces.obj), setting up the lighting and camera and rendering the image is shown in white, while the blue text shows the outputs from {\it Blender} during this process.  The 3D View panel (bottom middle) shows the resulting orientation of the surface and camera in the three dimensional space, while the UV Image Editor (top middle panel) shows the rendered image.  The Outliner panel lists the objects in the 3D View panel which include the loaded isodensity surfaces (surfaces), the camera (Camera) and the empty mesh used to control the camera's pointing (EmptyCamera).  }
\label{fig:simplerender}
\end{sidewaysfigure}

The pathway from an astrophysical dataset to a rendered image and/or three dimensional model revolves around first loading the data in a format that is understood by {\it Blender}, defining the lighting source(s), placing the objects in the three dimensional space, and then exporting a rendered image(s) or a three dimensional model.  The workflow of a {\it Blender} session is highly variable between users and projects.  Alternate workflows to that described here are demonstrated in the extensive introductions of \citep{kent2013,kent2015}.  For the purposes of this paper, the workflow will typically be composed of the six following steps, an example of which can be found in Figure \ref{fig:simplerender}.

\begin{itemize}

\item {\textbf{Load 3D Models}}: The first task is to load data files from one or multiple simulations.  One can either load pre-generated surfaces or particle clouds, or rely on the integrated {\it yt} \citep{turk2011} backend to generate these objects.  The {\cf Load} command in the {\it AstroBlend} library deciphers which type of file to import and is discussed more in section \ref{section:inputFormats}.  In the example code of Figure \ref{fig:simplerender} the file that is loaded is a set of OBJ-formatted surfaces from AMR data of an isolated galaxy simulation \citep{oshea2004} pre-generated with {\it yt}\footnote{See http://blog.yt-project.org/post/objexporter/ for more details on exporting OBJ-formatted surfaces}.  OBJ, or Wavefront, files are a data format used to represent objects in space as lists of the vertex locations, colors, and lighting options of individual polygons which, when combined, define a three dimensional surface.   
The surfaces stored in the OBJ file and loaded into {\it Blender} in Figure \ref{fig:simplerender} are two isodensity surfaces, colored by temperature.  In practice, one can load any series of surfaces in this manner including a time series.

\item {\textbf{Pick Lighting Source}}:  Currently, the supported sources of light are either {\it Blender}'s lamps which provide external sources of light, or the loaded objects themselves.  While point clouds which are used to represent particles in SPH data are generated with some emissivity, one can set the emissivity of a surface based on combinations of physical parameters used to extract the isosurface and its color map.  In Figure \ref{fig:simplerender}, the lighting is set to be solely due to the emissivity of the surface - in this example, it is a combination of the density (the isosurface value) and the temperature (isosurface color), $\epsilon \propto \rho^2 T^{1/2}$.  Note that for {\it Blender} to be able to import the emissivity of this externally generated OBJ file, slight modifications to {\it Blender}'s OBJ importer are necessary.  The few lines of code which need to be changed are discussed more fully on the AstroBlend website's ``Getting Started" page\footnote{\label{footn1}http://www.astroblend.com/getstarted.html}. 

\item {\textbf{Arrange the Scene}}:  Once objects are loaded and lighting sources are determined, one must place the camera, simulation outilizingbject and other three dimensional models in the scene.  In Figure \ref{fig:simplerender}, the three dimensional surfaces are automatically placed at the origin, and the camera, once initialized, is placed at 6~Blender Units (BU) along the negative x-axis, (-6, 0, 0), pointing at the center of our simulation object, (0,0,0).  More information about the camera setup is discussed in section \ref{section:cameraSetup}.  The lamp and cube from Figure \ref{fig:usual} are also deleted with the {\cf science.delete\_object} command.  One can move objects in the scene with commands in the Python console, or with a combination of the mouse and keyboard.  Here, direct interaction with the Python console is assumed and a full discussion of keyboard commands is relegated to the {\it AstroBlend} website\footref{footn1}.

\item {\textbf{Render an Image}}: A quick render to {\it Blender}'s Image Editor as depicted in Figure \ref{fig:usual} can be produced by pressing the {\cf camera} and {\cf Render} buttons, or render to a PNG image as shown at the end of the code segment in Figure \ref{fig:simplerender}.  To render to a file, {\it AstroBlend} requires a directory and a base naming structure for the image files and produces each rendered image tagged with a number.  Each time an image is rendered with this render object, the number tag of output image is increased by 1.  For example, the first render from the code in Figure \ref{fig:simplerender} would be ``galSurfs\_0000.png", and if {\cf render.render()} is called again, the next image would be stored as ``galSurfs\_0001.png".   This naming structure provides for easy combining of image files into movies externally from {\it Blender}.

\item {\textbf{Export 3D Objects}}: Before moving onto a new model, one can also export isosurfaces or save entire blend files to be later uploaded to various websites dedicated to the distribution of three dimensional models (see section \ref{section:sidebysideisos} for some examples).  

\item {\textbf{Delete the Model}}: When one is done with a model, it is important to fully remove the object from the 3D viewer and memory.  This is  accomplished by using the {\cf science.delete\_object(name\_or\_object)} where {\cf name\_or\_object} is either the name of the object, or the object variable itself.   As {\it Blender} is not designed for the rapid generation and removal of large polygon-count meshes, {\cf science.delete\_object} scripts delete the object from both the current 3D scene and memory as well (``unlinking" the object from the scene in standard {\it Blender} nomenclature).  
While this command is not explicitly shown for the isosurfaces in Figure \ref{fig:simplerender}, example usages of {\cf science.delete\_object} are shown in the removal of the cube and lamp objects in the arrangement of the scene.  This method should be utilized when scripting animations of large datasets.

\end{itemize}

It is worth noting that many of these operations can be done utilizing the GUI and mouse in addition to using {\it AstroBlend} in the Python Console\footnote{Several useful tutorials about setting up a test model and render using the GUI can be found here: https://en.wikibooks.org/wiki/Blender\_3D:\_Noob\_to\_Pro/Quickie\_Model and here: https://en.wikibooks.org/wiki/Blender\_3D:\_Noob\_to\_Pro/Quickie\_Render}.

\section{Data Import} \label{section:inputFormats}

In {\it AstroBlend}, all data files can be loaded with the same load command, {\cf science.Load}.  This command uses the file extension, or in the case of direct access to simulation files with {\it yt}, the {\it yt} simulation type flag, to determine how to load and display the model data.  Each source of model data has different parameters which can be used to augment what surface (for AMR data) or point cloud (SPH data) is displayed in {\it Blender}'s 3D viewer.

\subsection{Direct access with yt}

Currently, most of the widely used AMR codes supported by {\it yt} allow for isosurface models with {\it AstroBlend}, and many SPH codes supported by {\it yt} can be uploaded as point clouds.  
 {\it yt}, the widely used open source, parallel, visualization and analysis Python package, provides a wealth of tools to view and analyze data including volume rendering, projections, halo finding, isocontour generation, two dimensional histograms and integration of variables along user supplied paths.  At present {\it AstroBlend} mainly exploits {\it yt}'s ability as a data reader for both AMR and SPH codes, and its isocontour generation capabilities.  In addition, {\it AstroBlend} makes use of {\it yt}'s color maps and color mapping functions to rebin data values to one of the 256 indices in a specified {\it yt}  color map when generating surfaces or point clouds in {\it Blender}\footnote{More on {\it yt} color maps here: http://yt-project.org/doc/visualizing/colormaps/index.html}.

Table \ref{table:codesupport} lists the full range of code support available in both {\it yt} and {\it AstroBlend}.  For many simulation data types where {\it yt} does not support the generation of surfaces or point clouds, one can still load the files in {\it Blender}'s Python console, generate all supported {\it yt} analysis plots, including slice, projection and phase plots and save these files as pngs from the command line.  

As code support is being updated constantly, a more up to date list of supported codes can be found on the {\it AstroBlend} website\footnote{Up to date list of {\it yt} supported codes is in ``Current Code Support" section here: http://www.astroblend.com/tutorials/tutorial\_callingytdirectly.html}.  The parameters used in generating AMR surfaces and SPH point clouds determine the shape and color range of the displayed three dimensional models.

\subsubsection{Adaptive Mesh Refinement Data - Surfaces}

To upload an isosurface directly from an AMR simulation output file using {\it yt} as a backend, only slight modifications to  the {\cf science.Load} command from that listed in the code of Figure \ref{fig:simplerender} are necessary.  Specifically, one needs to pass information regarding which variable will be used to generate the isosurface, what variable to color the isosurface by, and how to make the surface emissive if the light source in the scene is not a lamp (more detail on lighting options in section \ref{section:lighting}).  Figure \ref{fig:ytamr} depicts several surfaces derived from an Enzo \citep{oshea2004} simulation along with their generating code.  

 In addition to specifying the isocontour parameter (density) and the color scheme (temperature), the call to {\cf Load} specifies the emissivity parameter of each face as a {\it yt} derived field from the combination of the isosurface generating and color parameters - in this example, proportional to the bremsstrahlung emission of the gas ($\epsilon \propto \rho^2 T^{1/2}$) in this example.  Here, {\it yt} is used to query the data, generate the isosurface vertices and color and emissivity maps, which are then passed to {\it Blender} which then generates the colored meshes in the three dimensional space from the vertex and color index lists.
Note that when generating isosurfaces directly with {\it yt} the physical parameters (scale, location, etc) of each surface can be manipulated independently, whereas in the case of the surfaces generated in {\it yt} externally from {\it Blender} of Figure \ref{fig:simplerender} all surfaces are loaded as a single mesh object.  However, if a Wavefront file generating program labels each surface independently in the OBJ file each surface can is uploaded as an independently named object.

\subsubsection{Smooth Particle Hydrodynamics Data - Point Clouds} \label{section:sphdata}

Direct access to SPH data is also available using {\it yt} as a backend in {\it Blender}.  An example calling sequence for generating a point cloud from {\it Tipsy} \citep{tipsy2011} simulation data along with a generated image is shown in Figure \ref{fig:ytsph}.  To create point clouds in a memory efficient manner, {\it AstroBlend} generates meshes with each vertex located at the position of an SPH particle.  By using a halo shader to create a point at the location of each vertex a point cloud can be rendered without the need to generate a sphere at each SPH location, a procedure which is highly memory intensive and slows render times considerably as discussed in \cite{kent2013}.
To simulate a range of point cloud colors, {\it AstroBlend} creates 256 separate meshes based on bins in the parameters {\cf color\_field} and {\cf color\_map}.
 In Figure \ref{fig:ytsph}, the point clouds are specified to be colored by the gas temperature using the Rainbow colormap.  Each vertex of each mesh is then rendered with the halo shader, with the size of the halo determined by the {\cf halo\_size} parameter.  The further away the camera is set from the point clouds, the larger the {\cf halo\_size} parameter needs to be to render each SPH particle.  
 
 In addition, while smaller halos in dense regions can be beneficial for visualization aesthetics, they can cause the gas to appear diffuse in less dense regions.  In future releases of {\it AstroBlend}, more freedom will be given to the user to modify the halo sizes of individual particles.
The related parameter {\cf clip\_begin}, shown in the example code of Figure \ref{fig:ytsph}, which determines the smallest object the camera will render can be lowered to insure both foreground and background SPH halos are adequately rendered.  Both this parameter and the {\cf halo\_size} parameter can be modified to optimize the final visualization.

\subsection{External File Formats} \label{section:externalFileFormats}

Beyond the use of {\it yt} as a backend for querying data and generating surfaces, {\it AstroBlend} supports two main external file formats for reading in simulation data - OBJ files and particle text files.  An example of reading in an OBJ surface pre-generated externally with {\it yt} was shown in Figure \ref{fig:simplerender}.  

Additionally, when SPH simulation data is not supported for direct access with {\it yt}, the output simulation data files can be converted to text particle lists for import with {\it AstroBlend}.  For this option, {\it AstroBlend} expects particle data in the following form: (column 1) the particle identifier (or a place holder if there is none), (columns 2-4) the x/y/z coordinates of the particle, (5) a number specifying the ``type" of particle.
For example, if one had formatted data from a simulation with gas particles (particle type 0) and star particles (particle type 1), the {\cf Load} command would be as follows: 
\begin{lstlisting}
myobject = science.Load(txtfile, scale = scale, 
	                halo_sizes = halo_sizes, particle_num=2, 
	                particle_colors=colors)
\end{lstlisting}
where {\cf txtfile} is the name of the particle text file, {\cf scale} is a tuple giving the x/y/z rescaling of the imported particle data set, {\cf halo\_sizes} gives the halo size of each of the {\cf particle\_num} particles, and {\cf particle\_colors} is a {\cf particle\_num} sized list of RGB tuples defining each particle group's color.
A more detailed description of importing data through these external formats is relegated to the online tutorials found on {\cf www.astroblend.com}.

\section{Lighting Options} \label{section:lighting}

Once AMR surfaces or SPH point clouds have been loaded into {\it Blender}, the source of light for the render must be selected.  Two options exist for illuminating scenes - the emissivity of the point cloud and surfaces themselves or external lamps.  
Both methods described here are meant to create photorealistic and not necessarily physically accurate images, and should be used for illustrative visualizations only.

\subsection{Available Lamps}

{\it Blender} provides several external lighting options to mimic spotlights, area lights, lamp light and more.  In {\it AstroBlend}, the currently support lamp types are {\cf POINT} and {\cf SUN}.  Both {\cf POINT}s, which are isotropic point light sources similar to the light produced from a naked light bulb, and {\cf SUN}s, which simulate far away point sources of light are initialized in the same way.  The following code snippet initializes a {\cf SUN} type lamp and places it at x/y/z coordinates (5,0,6) BU:
\begin{lstlisting}
light = science.Lighting(lighting_type=`SUN')
light.location = (5, 0, 6)
\end{lstlisting}
Other lamps not yet supported by {\it AstroBlend} can be added through the use of {\it Blender}'s GUI\footnote{See https://en.wikibooks.org/wiki/Blender\_3D:\_Noob\_to\_Pro/Understanding\_Blender\_Lights for a discussion of {\it Blender}'s many available lamps.}.

\subsection{Surface and Point Cloud Emission}

\begin{sidewaysfigure}
\centering
\includegraphics[width=1.0\textwidth]{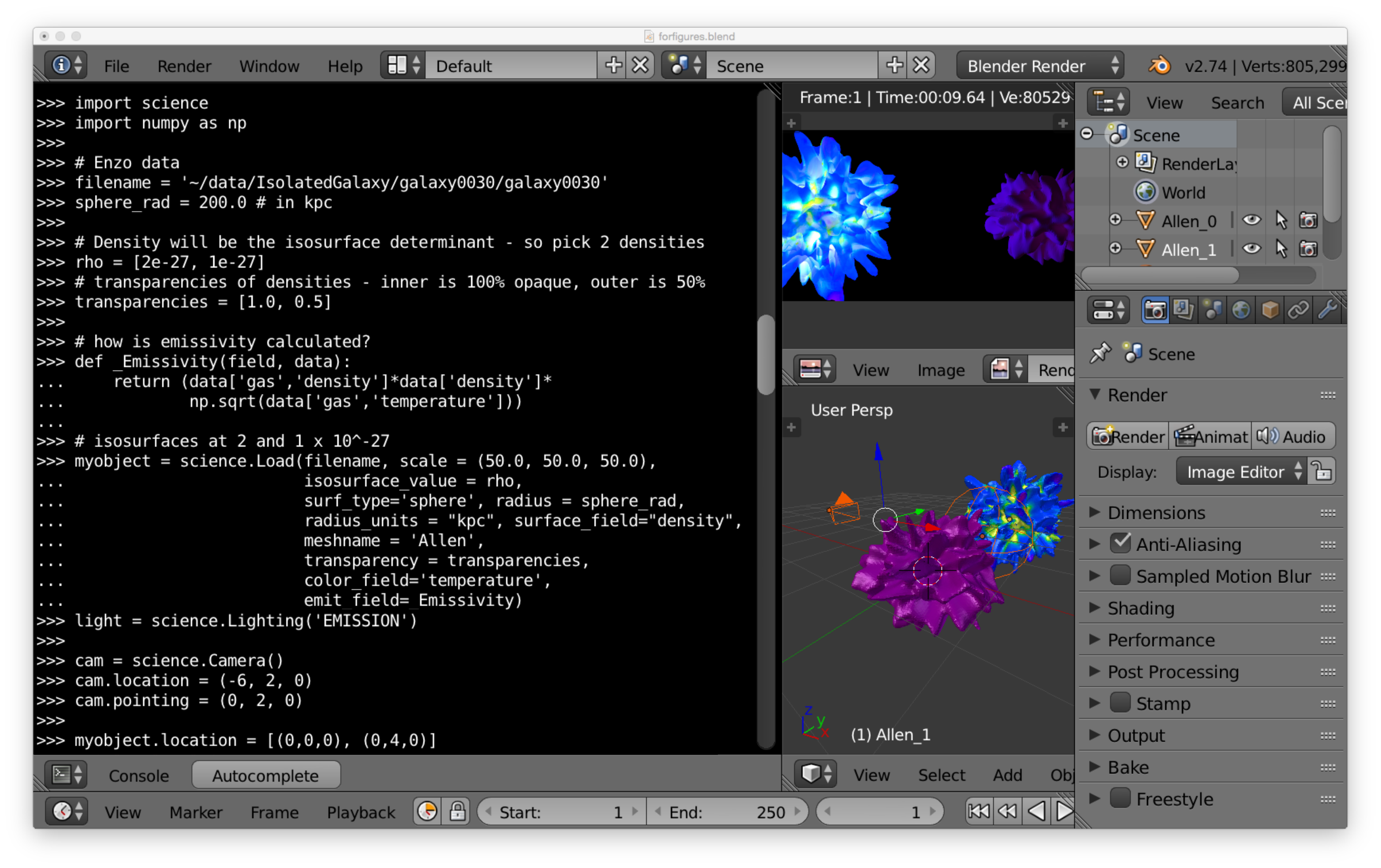}
\caption{Direct generation of isodensity surfaces from the isolated galaxy simulation of \cite{oshea2004} with {\it yt}.  Here, {\textit{yt}} \citep{turk2011} is called during the 
{\cf{Load}} command to create isocontours of the variable {\cf{surface\_field}} (density), at values given by {\cf rho} ($2 \times 10^{-27}$ and $10^{-27} \, {\rm g/cm^3}$), colored by {\cf{color\_field}} (temperature), with the emissivity of each face of the surface calculated by the function {\cf \_Emissivity} ($\epsilon \propto \rho^2 T^{1/2}$).  The code in the Python Console (left panel) results in an image similar to that shown in Figure \ref{fig:simplerender}, however, here the two isodensity surfaces are separated in the three dimensional space (bottom middle panel) for separate viewing in the final render (top middle panel) with the {\cf myobject.location} command. }
\label{fig:ytamr}
\end{sidewaysfigure}

\begin{figure*}
\centering
\includegraphics[width=0.75\textwidth]{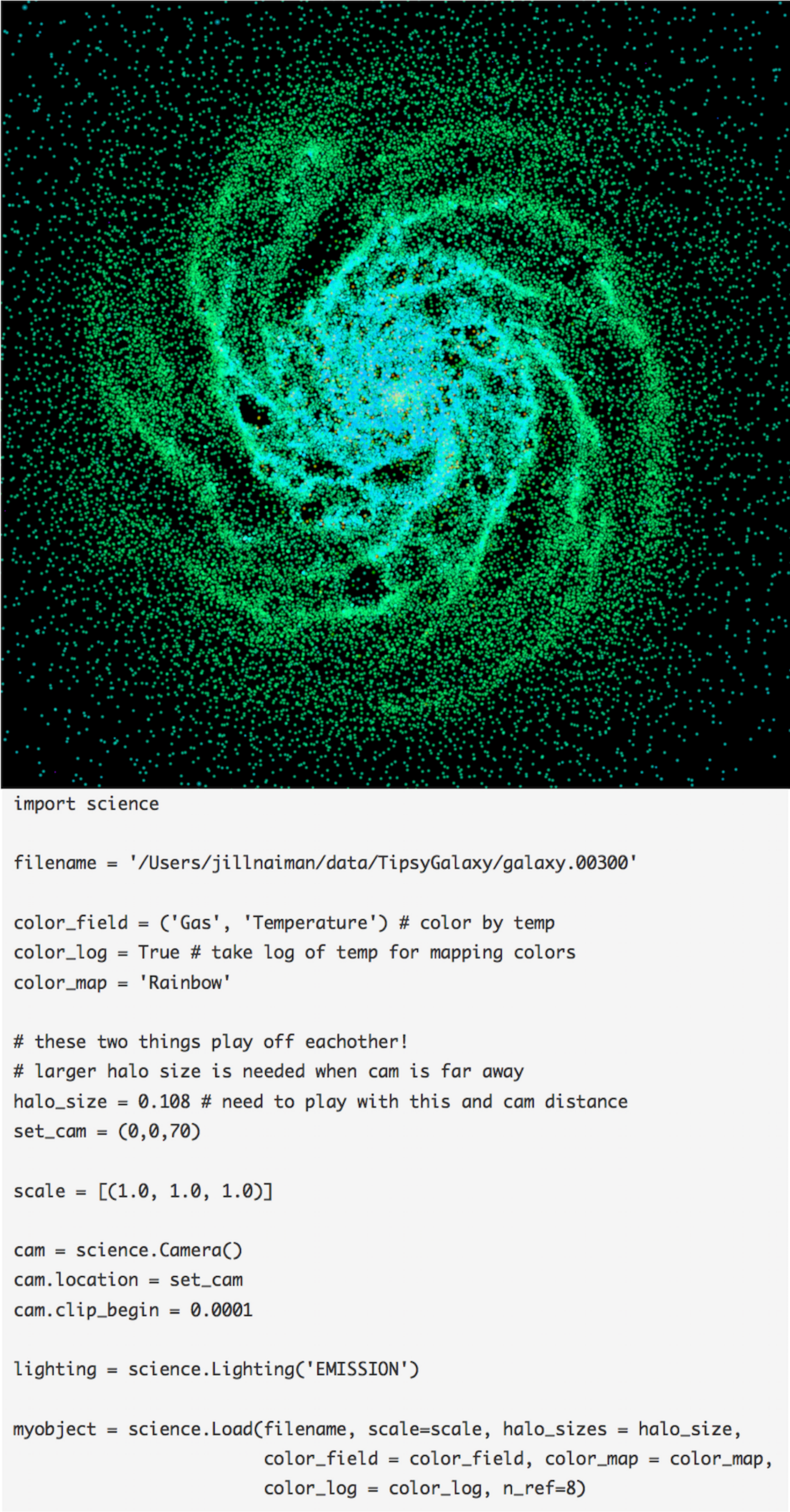}
\caption{The generation of a point cloud from the data of \cite{keller2014}.  Once again, {\it yt} is used to facilitate direct interaction with the Tipsy \citep{tipsy2011} dataset through the {\cf Load} command and its parameters.  Here, the different colors of the point cloud halo shaders are based on the temperature of each particle.  The camera is set above the galaxy along the z axis, and the clipping is lowered to be able to resolve the small halo shader particles.  The parameter {\cf n\_ref}, which controls the level of oct-tree refinement, is needed to use {\it yt} as a backend loader of this data. }
\label{fig:ytsph}
\end{figure*}

Emissive objects in {\it AstroBlend} come in two forms - glowing vertices from a halo shader for SPH point clouds (as depicted in Figure \ref{fig:ytsph}), and emission from the faces of an AMR isosurface (as shown in Figures \ref{fig:simplerender} and \ref{fig:ytamr}).  While halo shaders are emissive\footnote{For details on halo shaders in {\it Blender} see: https://www.blender.org/manual/render/blender\_render/materials/ special\_effects/halo.html} but do not cast light into the scene, an emissive isosurface can illuminate other objects in the scene.  In either case, emissive lighting must be selected ({\cf science.Lighting(lighting\_type = `EMISSION')}) in order to view the particles or surfaces in a render.

Mesh face emission is calculated with {\it Blender}'s implemented ``Approximate Ambient Occlusion" method.  In brief, this method approximates transmission and emission from different points on a mesh based on how much an emissive disk placed at each vertex would be occulted by other vertex's disks\footnote{More information on this method and theory papers found here: https://peach.blender.org/2008/01/approximate-ambient-occlusion/}.  While the level of emissivity of the SPH point clouds is fixed, each mesh face can have different emissivities based on their physical properties, as depicted in Figures \ref{fig:simplerender} and \ref{fig:ytamr}.  In Figure \ref{fig:simplerender} the emissivity of each surface is calculated externally in {\it yt} and read in with the other parameters of the OBJ file using the modified OBJ importer in {\it Blender} discussed in section \ref{section:workflow}.   When using {\it yt} within {\it Blender}, the emissivity of the sources is added as a derived field using {\it yt}'s standard prescription.  An example of this can be found in the code example of Figure \ref{fig:ytamr}.

\section{Camera Options} \label{section:cameraSetup}

\begin{figure*}
\centering
\includegraphics[width=0.75\textwidth]{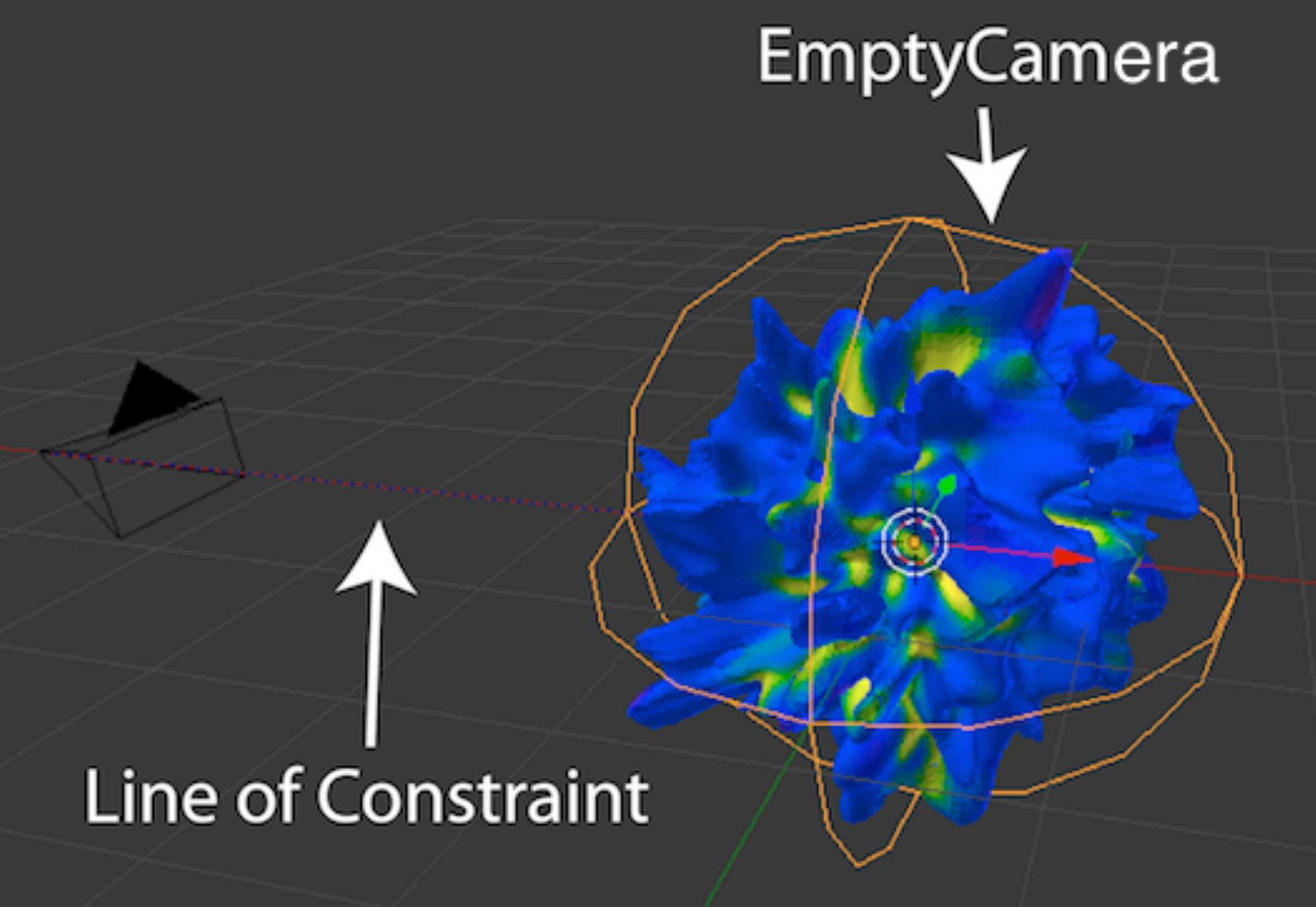}
\caption{Depicted is a close up of {\it Blender}'s 3D View containing a camera and its empty pointing mesh object as well as the isodensity surfaces generated in Figures \ref{fig:simplerender} and \ref{fig:ytamr}.  The pointing of the camera is parented to an empty mesh for ease of control and is initiated with a call of {\cf cam = science.Camera()}.  In this way, one can center the camera's pointing on an object - the isodensity surfaces from Figures \ref{fig:simplerender} and \ref{fig:ytamr} in this example.  The camera's location is modified by setting {\cf cam.location = (x,y,z)} and its pointing with 
{\cf cam.pointing = (px, py, pz)}.}
\label{fig:cameraTracking}
\end{figure*}

In addition to placing lamps and selecting sources of light, to render an image one must select a camera location and pointing.  {\it Blender} allows the user to move and rotate the camera directly in the 3D Viewer with the mouse\footnote{Many tutorials exist about manipulation of the camera.  One such example is http://wiki.blender.org/index.php/Doc:2.4/Manual/3D\_interaction/Navigating/Camera\_View}.  {\it AstroBlend} allows for manipulation of the Camera object and its pointing directly through the Python Console by parenting the Camera object to an ``Empty" mesh, as shown in Figure \ref{fig:cameraTracking}.
To initialize the camera with {\it AstroBlend} as shown in Figure \ref{fig:cameraTracking} the following single command is used:
\begin{lstlisting}
cam = science.Camera()
\end{lstlisting}
By parenting to a non-renderable mesh, the camera is easily pointed at an object by placing the empty mesh at the location of the object as shown in Figure \ref{fig:cameraTracking}, or by simply using the command:
\begin{lstlisting}
cam.location = (cam_x, cam_y, cam_z)
cam.pointing = (object_x, object_y, object_z)
\end{lstlisting}
where {\cf (cam\_x, cam\_y, cam\_z)} and {\cf (object\_x, object\_y, object\_z)} are the x/y/z coordinates of the camera and the object of interest, respectively.

\section{Three Dimensional Annotations} \label{section:annotations}

\begin{sidewaysfigure}
\centering
\includegraphics[width=1.0\textwidth]{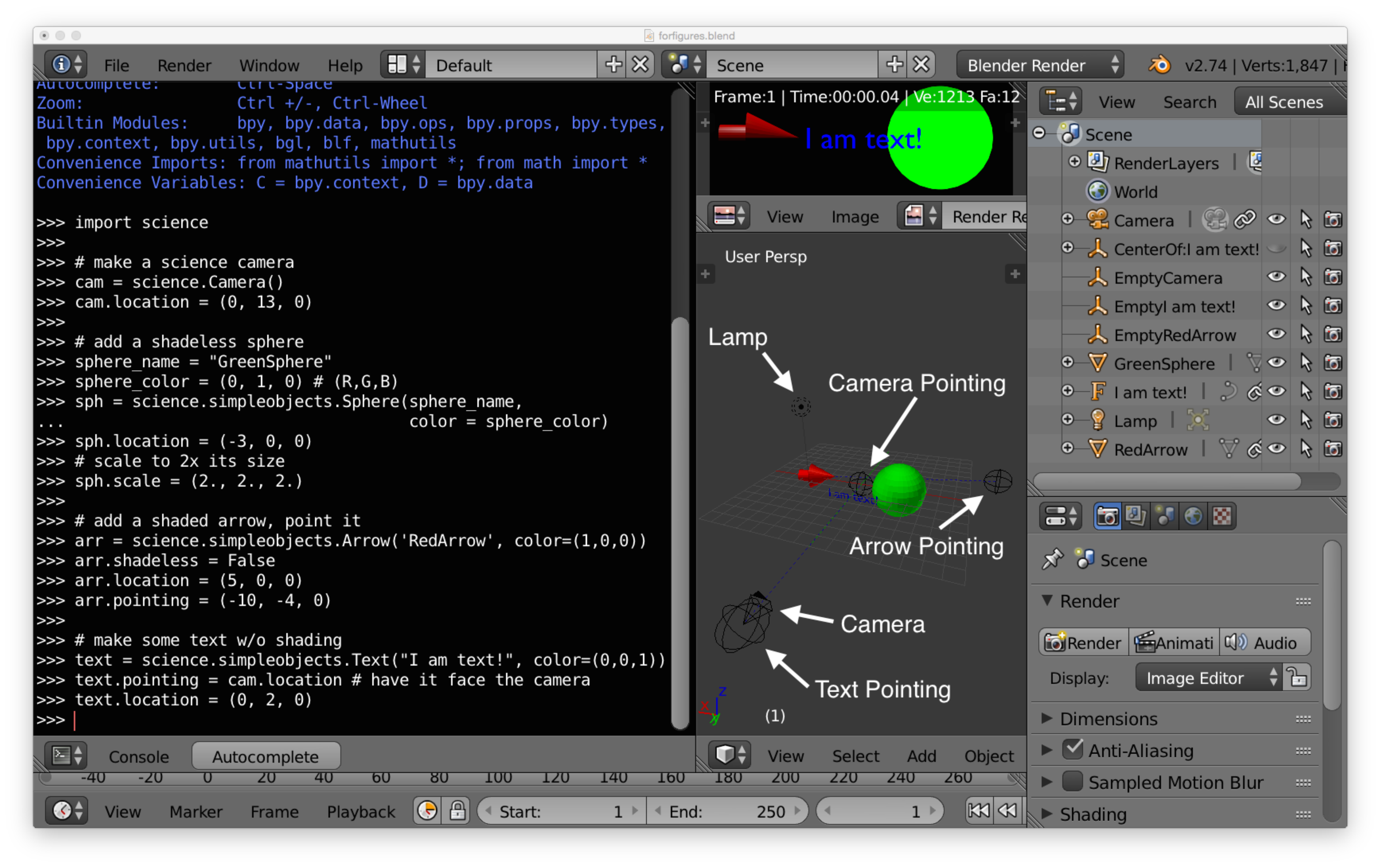}
\caption{Simple three dimensional objects are accessible through the python interface with the code shown in the Python Console (left panel).  The code snippet generates the three objects and various assorted empty meshes shown in the 3D View (bottom middle panel) and summarized in the Outliner (right upper panel).  The arrow object includes both the three dimensional model, and its empty pointing mesh labeled {\cf RedArrow} and {\cf EmptyRedArrow} in the Outliner, respectively.  The text object has three meshes associated with it - the blue text object ({\cf I am text!} in the Outliner), the empty pointing mesh ({\cf EmptyI am text!}) and the center empty mesh ({\cf CenterOf:I am text!}).  The text empty pointing mesh controls where the text is pointing while the center empty mesh controls the position of the center of the text - see text for further details. }
\label{fig:simpleobjects}
\end{sidewaysfigure}

Before rendering a final image consisting of loaded models, a placed camera and lights, the scene can be annotated with several simple three dimensional models.  {\it AstroBlend} currently supports annotations with colored spheres, arrows, and text.  While spheres are adirectional, pointings of both arrows and text are supported by the same method outlined in section \ref{section:cameraSetup} to enable camera pointing - these objects are parented to empty (non-renderable) meshes, and by moving these empty-meshes, the objects can be pointed.  Figure \ref{fig:simpleobjects} shows these three objects in {\it Blender}'s 3D viewer.  The code in the python console depicts the addition of a green sphere without shading (a sphere without depth in the render), an arrow with shading enabled, and a simple text statement.  All objects can be moved with their {\cf .location} parameters, and the text and arrow objects can be pointed in different directions with their {\cf .pointing} parameters.

Note in the top right Outliner panel of Figure \ref{fig:simpleobjects} there are three objects associated with the text object.  The {\cf CenterOf} object is an invisible empty-mesh at the center of the text object.  In order for the pointing empty-mesh to parent to the center of the text object, and not its leading side, an extra empty-mesh locked to the center of the text object is necessary.  In practice this requires the moving of any text to be performed by selecting the {\cf CenterOf} object, and not the text object itself.  More information on this topic can be found in the sixth tutorial on the {\it AstroBlend} website\footnote{http://www.astroblend.com/tutorials/tutorial\_simpleUseThings.html}.

\section{Complex Camera Motions for Movies} \label{section:cameraMotions}

The rendering of single images, either in {\it Blender}'s UV Image Viewer (as depicted with two button presses in Figure \ref{fig:usual}) or to a file (as shown in the code section of Figure \ref{fig:simplerender}) have been used as examples in the discussions of camera placement, lighting and annotation in the previous sections.  However, {\it AstroBlend} also supports the rendering of multiple images with a moving camera, which can be combined externally from {\it Blender} into a movie.  Currently supported camera motions include Zoom, Rotate and Bezier curves.

\subsection{Zoom}

The Zoom motion simply moves the camera along the line connecting the camera location to its pointing by a certain factor.  For example, to zoom the camera from 10 BU to 5 BU along the x axis one could use the sample code:
\begin{lstlisting}
render_directory = `/Users/jillnaiman/renders/'
render_name = `zooms_'
cam.location = (10, 0, 0)
cam.pointing = (0,0,0)
movie = science.Movies(cam, render_directory, render_name, 
                       render_type = `Zoom', render_steps = 60, 
                       zoom_factor = 0.5)
\end{lstlisting}
where {\cf render\_directory} is the directory where the renders will be stored and {\cf render\_name} is the base name of the renders.  The number of renders in the movie, given by {\cf render\_steps = 60} indicates there will be 60 files in the render directory with names from {\cf zooms\_000.png} to {\cf zooms\_059.png}.

\subsection{Rotation}

The Rotate camera motion rotates and zooms the camera about a fixed location.  The calling sequence is similar to the Zoom motion, but instead of specifying the factor by which to zoom, one specifies the rotation angles:
\begin{lstlisting}
cam.location = (10, 10, 10)
cam.pointing = (3, 0, 0)
render_name = `rotation_'
movie = science.Movies(cam, render_directory, render_name, 
                       render_type = `Rotation', render_steps = 60, 
                       radius_end = 2.0, theta_end = 0.1, 
                       phi_end = 90., use_current_start_angles=True, 
                       use_current_radius = True)
\end{lstlisting}
The above section of code rotates the camera around the point $(3, 0, 0)$ and zooms into a radius of 2.0 BU in 60 steps.  Each render is stored in {\cf render\_directory} as an image from {\cf rotation\_000.png} to {\cf rotation\_059.png}.  The keywords {\cf use\_current\_start\_angles} and {\cf use\_current\_radius} indicate that this movie will start with the current camera radius and rotation angles.  If these keywords are set to {\cf False} parameters defining the beginning radius and angles, {\cf radius\_start}, {\cf theta\_start} and {\cf pi\_start},  must be specified.

\subsection{Bezier Paths}

The final available camera motion, Bezier, interpolates the camera locations and pointings from given lists using bezier curves for a smooth, complex camera motion.  An example calling sequence uses three pointings and locations to constrain the camera motion along an interpolated 3D path:
\begin{lstlisting}
locs = [[10,10,10], 
	[10,-10,10], 
	[10,-10,-10]] 
pts =  [[0,3,0], [0,-3,0], [0,3,0]] 
movie = science.Movies(cam, render_directory, `bezier_', 
	               render_type = `Bezier', 
	               render_steps = 60, 
                       bezier_locations=locs, 
                       bezier_pointings=pts)
\end{lstlisting}
Using the above code, the camera will smoothly move thorough the points {\cf locs} while the camera's pointings are interpolated from the points {\cf pts}. \\
\\
A full discussion of these three camera motions with examples and movies is relegated to the seventh tutorial on the {\it AstroBlend} website\footnote{http://www.astroblend.com/tutorials/tutorial\_cameraMotions.html}.

\section{Example Visualizations} \label{section:examples}

The following section contains examples of {\it AstroBlend}'s visualization capabilities.  As these capabilities are constantly being extended, refer to the website's gallery and tutorials for further examples.
In all cases, references to timing are from the performance on a iMac quad core, 3.5 GHz Intel Core i7, 32 GB RAM desktop with a NVIDIA GeForce GTX 780M 4096 MB GPU. 

\subsection{SPH Galaxy Merger Simulation: A Simple Movie from Snapshot Files}

\begin{figure*}
\centering
\includegraphics[width=1.0\textwidth]{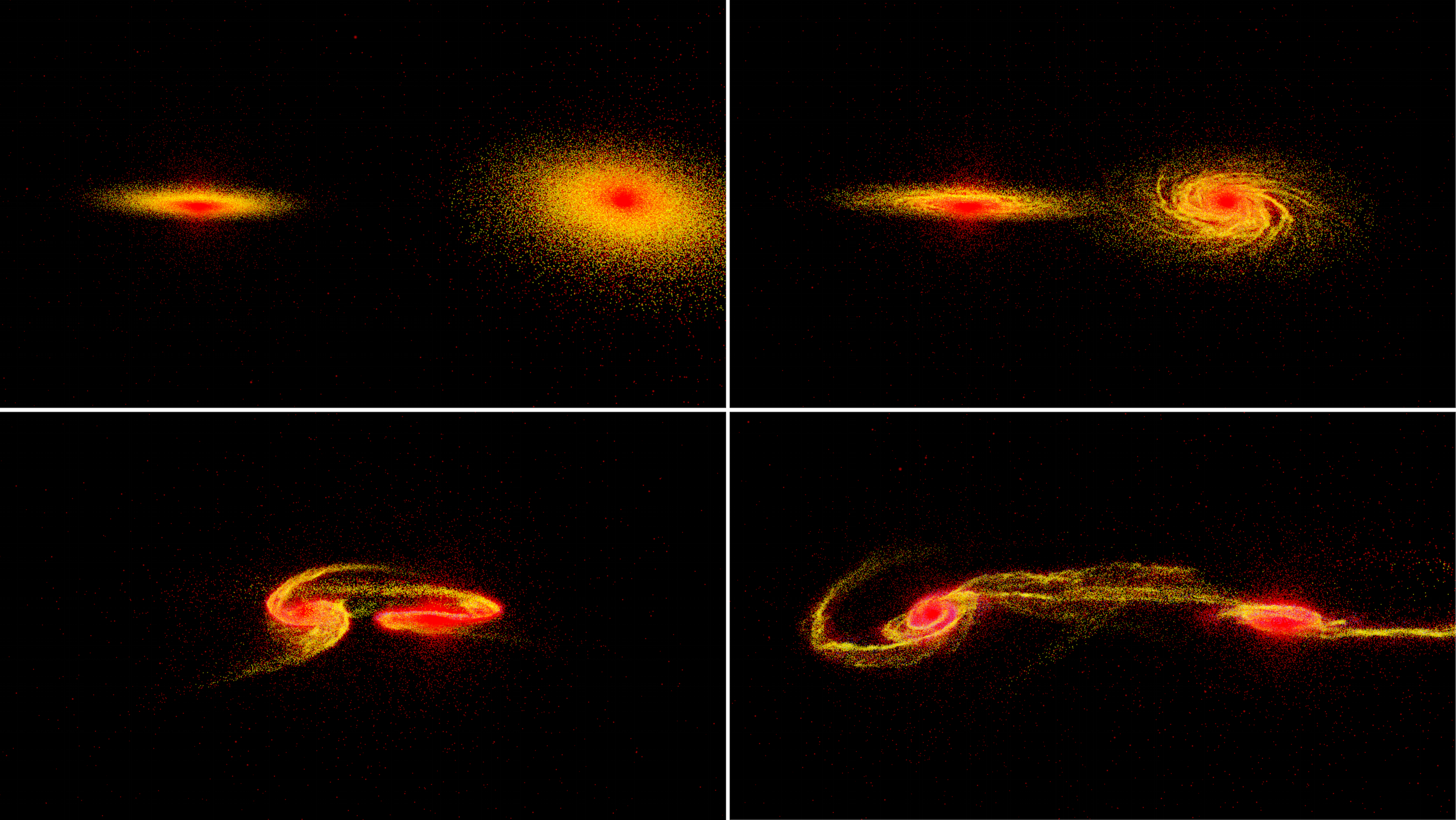}
\caption{Four renders from a movie generated in conjunction with the paper \cite{naiman2015} show different stages of a galaxy merger.  The particle colors represent pre-merger stars (red), gas (yellow), and newly formed stars (blue) as well as black hole sinks (green, not visible).  The full movie can be viewed at https://www.youtube.com/watch?v=8Snhd4B9SEc  }
\label{fig:sphmovie}
\end{figure*}

The snapshot files from the Gadget-2 \citep{gadget2} galaxy merger simulations of \cite{debuhr2011,debuhr2012} are rendered in series to generate a movie of the gas and stellar properties during a typical merger.  Figure \ref{fig:sphmovie} shows several frames of such a movie featured in \cite{naiman2015}.

Each render consists of approximately $1.6 \times 10^6$ particles, represented by the individual halo shaders as discussed in \ref{section:sphdata}.  Here, yellow particles are gas, red particles denote the original bulge and disk stars, and newly formed stars are blue.  The central massive black hole particles and dark matter halo particles are hidden from view in these movies to highlight the trajectories of the gas and stellar particles during the merger.  

The code used to generate the movie consists of first importing each sequential Gadget-2 snapshot file as a formatted particle text file with the {\cf Load} command and painting each particle with its specified color, determined by its particle type.  After an image is rendered to a PNG file, the particle cloud is deleted with {\cf science.delete\_object} before loading the next snapshot file.  In this example, the camera is stationary and the motion of the galaxies comes from the simulation itself. 

 This method of loading, rendering, and then deleting the data from each file serves to limit the memory usage of {\it Blender}, as apposed to loading the full set of datafiles to memory and animating their visibility in time.  An individual cycle of loading a data snapshot and rendering an image takes $< 20$~s.  However, repeated loading and rendering of data and images results in a slowdown of each loading processes such that the resultant movie in this example movie took approximately 7 hours to generate.  This bug is likely due to imperfect memory reallocation and will be addressed in a future release of {\it AstroBlend}.

A full, high resolution, movie of this simulation is available on the {\it AstroBlend} Youtube site\footnote{https://youtu.be/8Snhd4B9SEc}.

\subsection{Side by Side Isosurfaces: Export Data to A Variety of 3D Websites} \label{section:sidebysideisos}

\begin{figure*}
\centering
\includegraphics[width=1.0\textwidth]{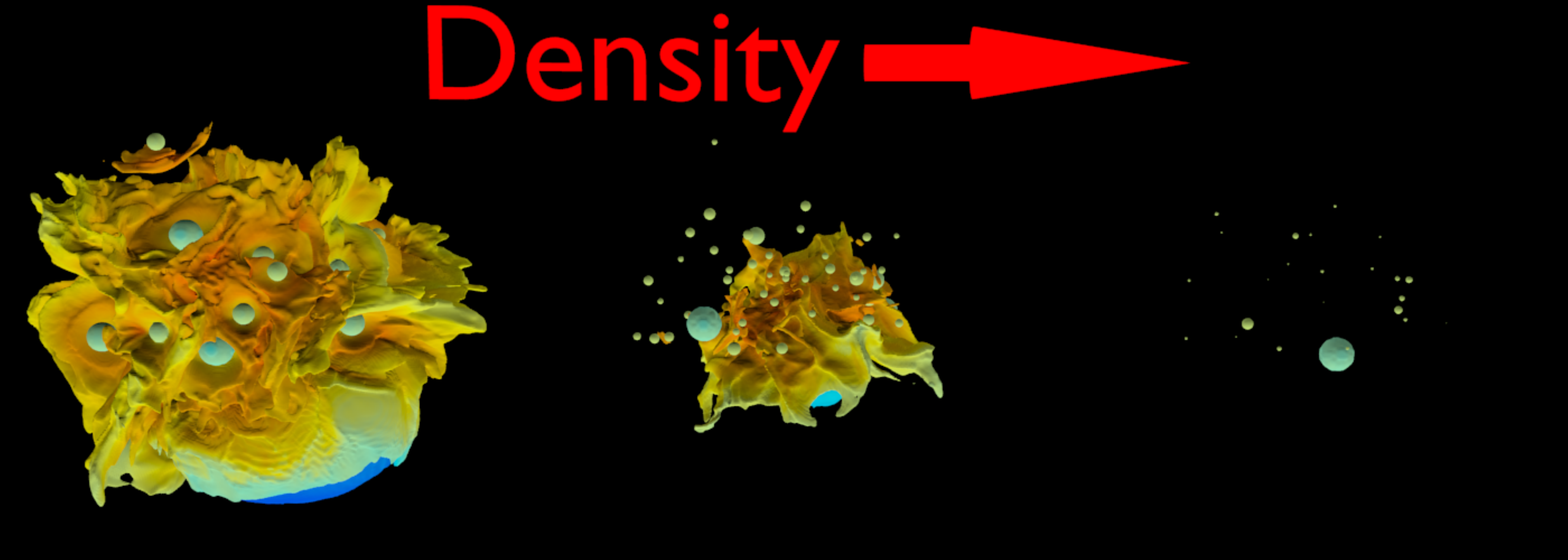}
\caption{Several isodensity contours created in conjunction with (Soares-Furtado et.\ al, in prep.) show the gas structures surrounding stellar winds in a present day M15 globular cluster analog.  The surfaces are colored by temperature with the cold gas residing in the areas directly surrounding the stars (blue) and the hot gas forming in the shocked colliding stellar winds (red).  As the density of each isocontour increases (from left to right), we are probing the structures which closely surround each individual star.  This model is available interactively here: https://skfb.ly/IwyK.}
\label{fig:sidebyside}
\end{figure*}

In addition to producing movies from simulation snapshot files, several three dimensional data products can be formatted in {\it Blender} for 
export to a variety of new model sharing websites.  Figure \ref{fig:sidebyside} shows a rendered image of one such data product - three isosurfaces of decreasing density, colored by temperature.  This figure is a modified version of Figure 3 in Soares-Furtado et al.\ (in prep) which depicts the gas structure created by the evolved stellar population in the present day globular cluster M15, representing $\approx 10^{8}$ cells of a Cartesian geometry FLASH \citep{fryxell2000} simulation.

While Figure \ref{fig:sidebyside} illustrates an essential figure in a stellar winds paper (Soares-Furtado et al., in prep), this particular visualization also exemplifies how to make visualizations accessible across various forms of media.  As an example, these isosurfaces and annotations have been published to the Sketchfab website allowing for realtime interaction with the model in three dimensions \footnote{Main Sketchfab website: https://sketchfab.com/, specific winds model: https://skfb.ly/IwyK}.  Sketchfab allows for the publication and distribution of interactive models through embedded interfaces on a website or blog and downloadable OBJ files for other users.  To share an isosurface with collaborators or the general public, one can directly upload {\it Blender} files, or export OBJ files using the {\cf science.scienceutils.export\_obj} command.  While the loading of the data in this example is relatively fast (1-30~seconds for each isosurface), the exporting as an OBJ file took longer than rendering (10~seconds to render, 1~minute to export).
More details about this example can be found in the fourth tutorial on {\cf www.astroblend.com}\footnote{More information about direct interaction with data through {\it yt} can be found on http://www.astroblend.com/tutorials/tutorial\_callingytdirectly.html}.

\begin{figure*}
\centering
\includegraphics[width=1.0\textwidth]{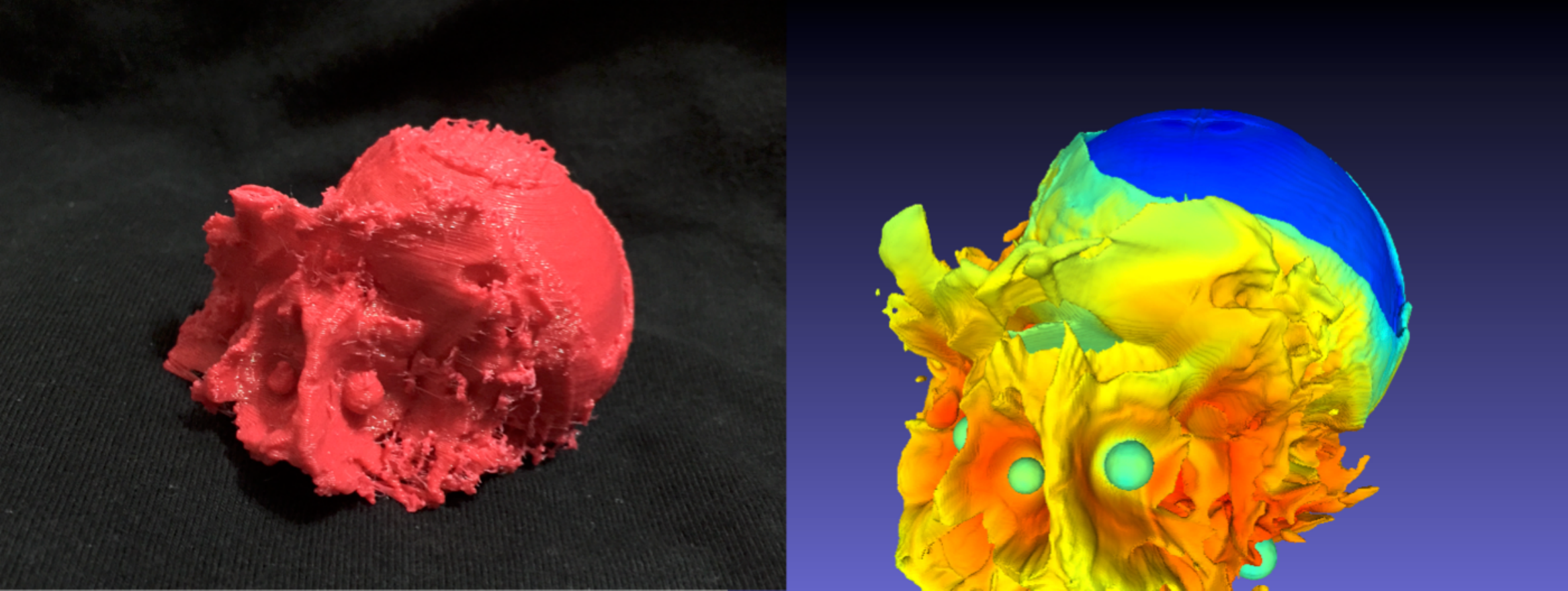}
\caption{Isocontours generated with {\it AstroBlend} can be used in 3D printing.  The isodensity contour of the stellar wind structures modeled in Soares-Furtado et.\ al, (in prep.) on the right is available as a 3D printable stl file as shown on the left, on Thingiverse: http://www.thingiverse.com/thing:1145716.}
\label{fig:3dprint}
\end{figure*}

The largest of the three isosurfaces depicted in Figure \ref{fig:sidebyside} has also been made available as a 3D-printable file on Thingiverse\footnote{Main Thingiverse website: https://www.thingiverse.com/, specific winds model: http://www.thingiverse.com/thing:1145716}, a repository for freely available three dimensional printing model files.  OBJ files exported from {\it Blender} can be uploaded into many open source 3D printing software packages, which can then be configured to print on specific 3D printers \citep{steffen2014,vogt2014,madura2015}.  In Figure \ref{fig:3dprint}, the original isosurface is shown alongside a photograph of its 3D-printed version.   Uploading isosurfaces to 3D-printing distribution sites allows the scientist to print their models themselves, or make them available to the public for artistic or educational purposes, as exemplified by the use of such surfaces to make ``visualizations" for the blind\footnote{Downloadable 3D printable models for the blind: http://3dprint.com/72633/think3d-devnar-for-the-blind/}.

\subsection{Including Artistic Models in your Visualization}

As {\it Blender} is first and foremost a 3D modeling and graphics software package, this last example visualization touches briefly on {\it AstroBlend}'s capabilities to combine three dimensional artistic models with astrophysical data and simulation snapshot files.  

\begin{figure*}
\centering
\includegraphics[width=1.0\textwidth]{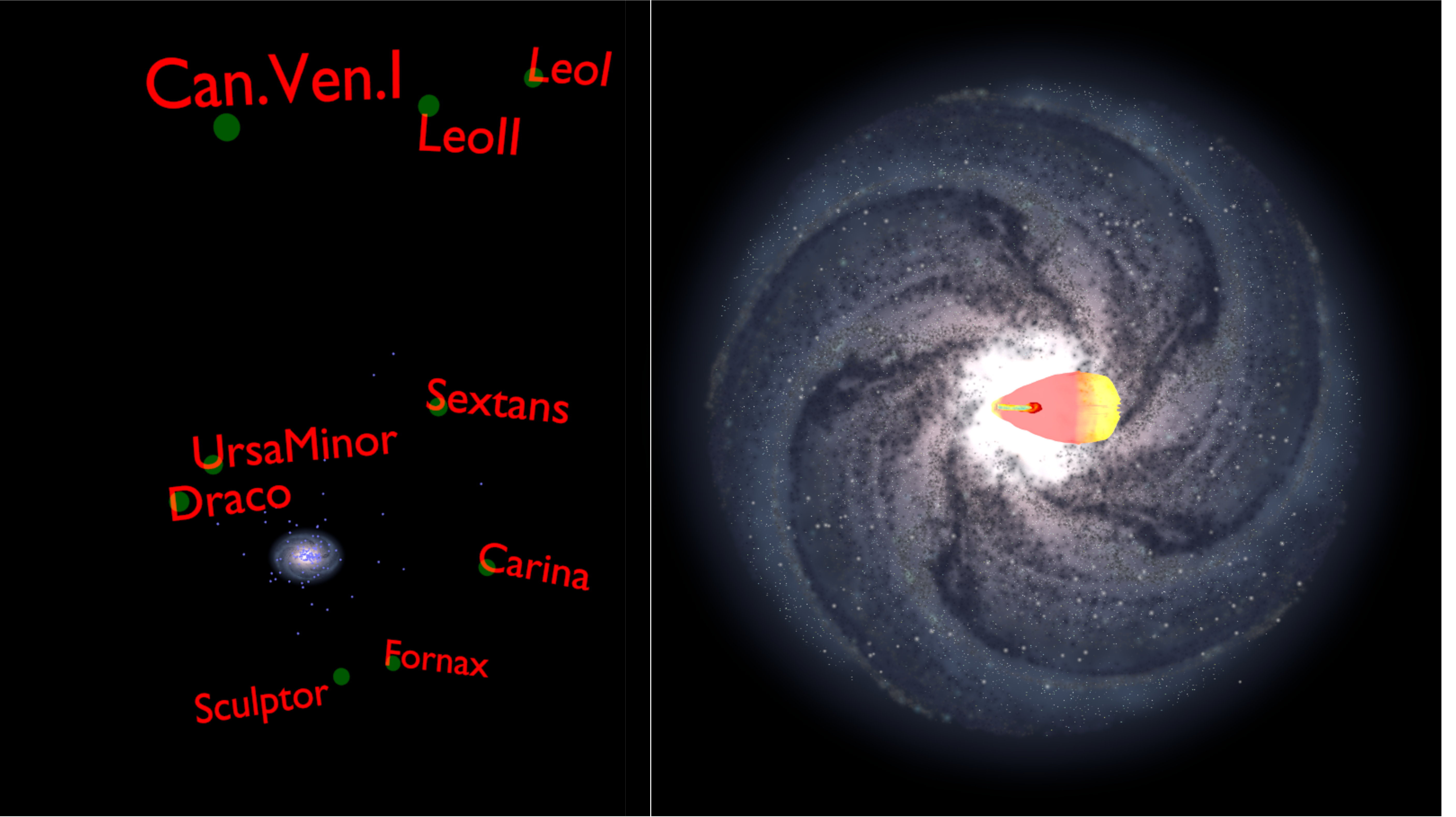}
\caption{The use of an artistic three dimensional model alongside astrophysical data can create context for simulation and observational data. On the right a model galaxy is combined with the observed positions of Milky Way globular clusters (small blue spheres) from \cite{harris1996} and dwarf galaxies (labeled, larger green spheres) tabulated in \cite{deason2011}. A modified version of the right hand panel can be found on Sketchfab: https://skfb.ly/IqAF.}
\label{fig:art}
\end{figure*}

Figure \ref{fig:art} shows how one can combine an artistic three dimensional model with scientific data.  Here, the user can download a pre-made model of a physical object as a {\it Blender} file and upload text or simulation data using the Python API and {\it AstroBlend}.  In the right panel of Figure \ref{fig:art}, the positions and names of several Milky Way dwarf galaxies are shown with an artistic model of the Milky Way, to scale.  This model Milky Way is part of a downloadable Youtube tutorial on making composite objects with particles and vortices in {\it Blender}\footnote{\label{footn2}Model from the YouTube tutorial (https://youtu.be/8aVOQw-BIf0) of Thomas Piemontese, tomwalks.deviantart.com}. A modified version of this model can be viewed interactively on Sketchfab as well\footnote{https://skfb.ly/IqAF}.  The left panel of Figure \ref{fig:art} shows the model galaxy and the bowshock around a simulated dwarf galaxy as it is ram pressured stripped during its incorporation in the Milky Way's halo.  Here, the artistic galaxy is used as a point of reference for the size of the dwarf galaxy and its bowshock.  Both of these images took less than 5 seconds to load and render.

By using this three dimensional artistic model of a galaxy, one can quickly understand the scale of observed substructures in our Galaxy or place simulation data done in a small scale box into the larger astrophysical context of the Milky Way.

\section{Summary and Future Plans} \label{section:discussion}

As simulated and observed datasets increase in size, scientists need improved tools to analyze and visualize multidimensional datasets.  In this work I introduce {\it AstroBlend}, a Python library for use in the 3D animation software {\it Blender}. {\it AstroBlend} relies on the isosurface mesh creation, point cloud generate, complex camera motions, lighting and texturing available in {\it Blender} to visualize both AMR and SPH simulation data.  The use of {\it yt} as a backend in {\it AstroBlend} to generate surface and particle visualizations from direct interaction with simulation data files is discussed.
In the final section of this paper, several examples of the creation of movies, three dimensional data projects, and the integration of 3D artistic models with astrophysical datasets are discussed.  These examples serve as a basis with which the reader can begin to explore options for generating their own interactive visualizations for papers and press releases.

 Future versions of {\it AstroBlend} will permit the user to do on-the-fly data analysis using {\it yt} and make further use of {\it Blender}'s volume rendering capabilities and external rendering engines.  Currently the development version supports a GUI for {\it AstroBlend} features and the display of {\it yt} analysis plots within {\it Blender} and some preliminary volume rendering.  Once these updates have been pushed to the stable version of {\it AstroBlend}, further future updates will include the Cycles rendering engine and Cycles point density support for SPH simulations.
 
 The reader is encouraged to visit the AstroBlend website {\cf www.astroblend.com} for further information, resources, and tutorials.\\
\\
The author would like to acknowledge illuminating discussions with Matthew Turk, Morgan MacLeod, Melinda Soares-Furtado and Enrico Ramirez-Ruiz.  In addition, the thorough comments of one anonymous reviewer and the extensive input of another, Dr. Rhys Taylor were invaluable.  This work is supported by an NSF grant AST-1402480.

\clearpage

\begin{table}[htb!]
\footnotesize
  \caption{{\it AstroBlend} Code Support for Direct Data Access with {\it yt}}
  \label{table:codesupport}
  \begin{center}
\renewcommand{\arraystretch}{1.3}
    \begin{tabular}{clll}
    \hline
Code & Code/Format Type & AB Surface Support & AB Point Cloud Support  \\ \hline\hline
FLASH  & AMR & Y & NA \\
Enzo & AMR & Y & NA \\
Athena & AMR & Y & NA \\
Artio & AMR & N\textsuperscript{1}  & NA \\
Fits & FITS & N\textsuperscript{2}  & NA \\
GDF & Grid Data & N\textsuperscript{2}  & NA \\
MOAB & AMR & N (partial loading)\textsuperscript{2} & NA \\
SPH Text Files & Text Formatting & NA & Y\textsuperscript{1} \\
Tipsy & SPH & NA & Y \\
Gadget & SPH & NA & Y  \\
 \hline
\multicolumn{4}{l}{\textsuperscript{1}\footnotesize{See section \ref{section:externalFileFormats} for more details.}}\\
\multicolumn{4}{l}{\textsuperscript{2}\footnotesize{One can load and make plots in {\it Blender} via yt with this simulation type.}}\\
    \end{tabular}
  \end{center}
\end{table}

\bibliographystyle{elsarticle-num}

\bibliography{bib_2015}

\end{document}